\documentstyle[prl,aps]{revtex}
\input epsf.sty
\begin{document}
\draft
\draft
\twocolumn[\columnwidth\textwidth\csname@twocolumnfalse\endcsname
\title{Microscopic structure of fundamental excitations in $N$=$Z$ nuclei}
\author{Wojciech Satu{\l}a$^{(1,2)}$ and Ramon Wyss$^{(1)}$}
\address{$^{(1)}$ Royal Institute of Technology, Frescativ. 24,
S-104 05 Stockholm, Sweden}
\address{$^{(2)}$Institute of Theoretical Physics, Warsaw University,
ul. Ho\.za 69, PL-00 681, Warsaw, Poland }
\date{\today}
\maketitle
\begin{abstract}
Excitation energies of the $T$=1 states in even-even 
as well as  $T$=0 and $T$=1 states 
in odd-odd $N$=$Z$ nuclei are calculated within the mean-field approach.
It is shown that the underlying structure of these states 
can be determined in a consistent manner
only when both isoscalar and isovector pairing collectivity
as well as isospin projection, treated within the iso-cranking 
approximation, are taken into account. In particular, in
odd-odd $N$=$Z$ nuclei, the interplay between quasiparticle excitations
(relevant for the  case of $T$=0 states) and iso-rotations 
(relevant for the case of $T$=1 states) explains the near-degeneracy
of these fundamental excitations.
\end{abstract}
\addvspace{5mm}]
\narrowtext

%---------------------introduction---------------------
It is well known that pairing properties of finite Fermi systems
are number-parity dependent.
This is particularly well documented in atomic nuclei which
exhibit phenomena like odd-even mass staggering or odd-even staggering
of the moments of inertia.
These phenomena origin from simple phase-space
quenching due to the odd (quasi)particle known as the blocking effect.
Within the standard BCS theory of superconductivity the blocking effect
can be naturally accounted for by assuming the ground state of
the odd system to be a one-quasiparticle (qp)
[or two-quasiparticle (2qp) in odd-odd (o--o) nuclei]
excitation on top of the even-even vacuum, $\alpha^\dagger |vac\rangle$.
In fact, the simplicity and consistency of the BCS treatment of
even and odd nuclei was of paramount importance to establish the
theory of superconductivity in atomic nuclei~\cite{[Boh58],[Bel59]}.

The classical BCS theory
requires to be extended only in the closest vicinity of 
the $N$=$Z$ line. 
In these nuclei, apart from the isovector
pairing mode, also isoscalar neutron-proton Cooper pairs coupled to
non-zero angular momentum can be formed. However, the
empirical fingerprints of this pairing phase are
not very clear. The problem is rather complex, because it
requires a detailed understanding of both pairing phenomena
and the nuclear symmetry energy.
An invaluable source of information allowing to disentangle
these effects, are the isobaric
excitations in $N$=$Z$ nuclei as already discussed
in~\cite{[Jan65],[Zel76],[Vog00],[Mac00s]}.
Unfortunately, most of these studies were either purely
phenomenological or based on, in our opinion, inconsistent  models. 
In this letter,
we argue that the proper understanding of the isobaric excitations 
can be obtained only on a microscopic level.
It requires that both
isoscalar and isovector pairing as well as
isospin projection [at least approximate] 
are taken into account.
Moreover, within such a 
model, the standard BCS scheme of elementary 
excitations does not apply any longer. 
We provide the necessary extensions of the BCS theory which 
allow for a simultaneous description of 
{\bf (i)} the mass-excess in $N$=$Z$ nuclei,
{\bf (ii)} isospin $T$=1 excitations
in even-even (e--e) $N$=$Z$ nuclei [theory of  $T$=2 states in e--e
nuclei was given in our previous letter~\cite{[Sat00b]}], and
{\bf (iii)} $T$=0 and $T$=1 states in o--o nuclei.

%----------------- T=1 in even-even - sp model ----------------------
%%%%%%%%%%%%%%%%%%%%%%%%%% figure 1 : sp-model %%%%%%%%%%%%%%%%%%%%%%%%%%
\begin{figure}
\begin{center}
\leavevmode
\epsfysize=8.0cm
\epsfbox{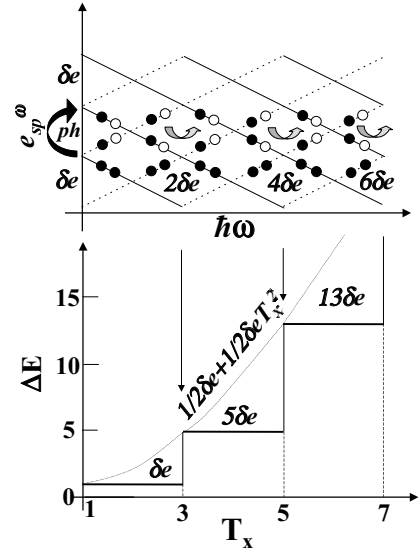}
\end{center}
  \caption{The single-particle routhians (upper panel) versus
  iso-cranking frequency for the equidistant level model. Solid (dashed)
  lines depict downsloping, $|+\rangle$, and upsloping, $|-\rangle$,
  sp states carrying iso-alignments of $\pm 1/2$, respectively.
  At each level crossing (indicated by arrows) the configuration changes,
  and hence excitation energy and iso-alignment (lower panel).
  }
  \label{fig1}
  \end{figure}
%%%%%%%%%%%%%%%%%%%%%%%%%%%%%%%%%%%%%%%%%%%%%%%%%%%%%%%%%%%%%%%%%%%%%%%%

We begin with the description of the $T$=1 states in e--e nuclei.
Similar to our letter~\cite{[Sat00b]} we start with
a  single-particle (sp) model, 
$\hat H^\omega=\hat H_{sp}-\hbar\omega \hat t_x$,
where $\hat H_{sp}$ generates (for a sake of simplicity) an equidistant
and iso-symmetric i.e. four-fold degenerate spectrum $e_i=i\delta e$.
For non-zero frequency each level splits into a pair of 
upsloping ($|-\rangle$)
and downsloping ($|+\rangle$) routhians, 
carrying alignment $t_x = \mp 1/2$.
The sp routhians cross at the frequencies:
$\hbar\omega_c^{(n)} =  (2n-1)\delta e$ where $n=1,2,3...$.
As shown in~\cite{[Sat00b]} the reoccupation process which takes
place at each level crossing [$\omega_c^{(n)}$] 
conserves the iso-signature symmetry.
In other words, cranking the lowest sp configuration (vacuum) gives
only states of {\sl even} isospin.
Hence, states of odd isospin are
obtained by promoting one particle
from the $|-\rangle$ state to the lowest $|+\rangle$ state as
depicted in Fig.~\ref{fig1}. 
The lowest odd-T branch of the iso-rotational band is obtained
by cranking this particle-hole
(p-h) excited state. The excitation energy and initial alignment
of this p-h state are $\delta e$ and $T_x$=1, respectively.
The p-h excitation modifies the iso-rotational spectrum in the
following manner:
First of all it blocks the level crossing at $\hbar\omega = \delta e$.
The allowed crossings appear at frequencies $\hbar\omega_c^{(n)}=2n\delta e$
[$n$=1,2,3...]. At each level crossing
the isospin changes by $\Delta T_x$=2 giving rise
to an odd-T iso-rotational band. The total excitation energy
of the band follows
\begin{equation}\label{spe}
\Delta E = {1\over 2}\delta e + {1\over 2}\delta e T_x^2
\end{equation}
dependence. The formula (\ref{spe}) is similar to the one obtained
previously~\cite{[Sat00b]} for
the states of even isospin. Indeed,
the moment of inertia (MoI) of the 
even-T iso-rotational band [built on the
vacuum] comes out identical to the MoI of the odd-T band [built on
the lowest p-h excitation]. However, the odd-T band is shifted in energy
by $\delta e$/2 due to the required p-h excitation.
Although the sp model is oversimplified, it reveals
the nature of odd- and even-T states
in e--e nuclei:
Due to iso-signature symmetry, odd-T states are based
on an excited p-h configuration and 
cannot be reached by iso-cranking the 
vacuum configuration to an odd-T$_x$ value. 
For a realistic model
including pairing correlations,
this corresponds to the lowest two-quasiparticle (2qp) state.

%----------  WS lipkin nogami calcs of e--e nuclei

We now proceed to investigate how
the excitation scheme is modified in the presence of
pairing correlations. Our hamiltonian is based on the deformed
single particle potential of Woods-Saxon (WS) type~\cite{[Cwi87s]}.
The two-body correlations contain both isovector and
isoscalar seniority-type pairing:
\begin{equation}
\hat H^\omega = \hat h_{WS} +G_{T=1}\hat P_{1}^\dagger
\hat P_{1} +
G_{T=0} \hat P_{0}^\dagger \hat P_{0}
-  \hbar \omega \hat t_x
\label{ham}
\end{equation}
where $\hat P_{1}^\dagger$ and $\hat P_{0}^\dagger$
create isovector and isoscalar pairs, respectively.
The Hamiltonian (\ref{ham}) is solved using the Lipkin-Nogami
method. 
The model is very similar to the one described 
in detail in Ref.~\cite{[Sat00a]}. However, different to
Ref.~\cite{[Sat00a]} we now employ
the most general Bogolyubov transformation. It allows us to
fully explore the isoscalar pairing channel without
any symmetry induced restrictions
i.e. to include
simultaneously {\boldmath{$\alpha\alpha$}} and
{\boldmath{$\alpha\tilde\alpha$}}
isoscalar pairs.
It is important to stress that the model is identical the one used 
in Ref.~\cite{[Sat00b]} to compute $T$=2 excitations in e--e nuclei. 
Like before, we set the deformation
to $\beta_2=0.05$. In the applications to e--e nuclei we use the 
same values of $x^{T=0}=G_{T=0}/G_{T=1}$ and cut-off parameters,
while for the o--o ($N$=$Z$=$A$/2) cases the 
cut-off parameters and $x^{T=0}$ values deduced for the e--e
$N$=$Z$=$A$/2--1 neighbour were used consequently.
Excitation energies as discussed in this work are
not affected by the kind of $T$=0 pairing used in the
calculations ({\boldmath{$\alpha\tilde\alpha$}},
{\boldmath{$\alpha\alpha$}} or mixed phases). In the following we
restrict the calculations to ({\boldmath{$\alpha\tilde\alpha$}}
pairing in the $T$=0 channel, which in the presence
of standard cranking corresponds to states of low angular 
momentum~\cite{[Sat00a]}.

%%%%%%%%%%%%%%%%%%%%%%%%%% figure 2 : T1-excitations %%%%%%%%%%%%%%%%%%%%%%%%%%
\begin{figure}
\begin{center}
\leavevmode
\epsfysize=5.0cm
\epsfbox{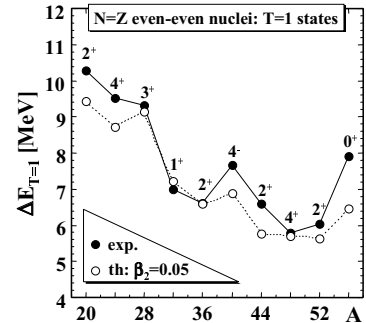}
\end{center}
  \caption{Excitation energies, 
   $\Delta E_{T=1}$, and angular momenta
   $I^\pi$ of the lowest $T$=1 
   states in e--e nuclei ($\bullet$). 
   The calculated results are marked by ($\circ$).}
  \label{fig2}
  \end{figure}
%%%%%%%%%%%%%%%%%%%%%%%%%%%%%%%%%%%%%%%%%%%%%%%%%%%%%%%%%%%%%%%%%%%%%%%%

In our previous letter we have shown that the cranked ground
state configuration $|vac\rangle$ with
$T_x$=$\sqrt6$ [$T_x$=$\sqrt{T(T+1)}$]
yields surprisingly accurate predictions for $\Delta E_{T=2}$.
However, as mentioned above we cannot repeat this
procedure for the $T$=1 states and determine the
frequency $\hbar\omega$ for the
$|vac\rangle$ so that $T_x$=$\sqrt{2}$.
This would violate the iso-signature symmetry. 
The proper trial wave function corresponds to
the lowest 
elementary excitation, i.e. to 
2qp, $\hat \alpha_1^\dagger \hat\alpha_2^\dagger |vac\rangle$,
excitation.
Hence, we do proceed as follow: ({\it i\/}) At each 
iteration step we perform the standard Hartree-Fock-Bogolyubov (HFB) 
transformation~\cite{[RS80]}
\begin{equation}
\left(
\begin{array}{c}
 \mbox{{\bf U}}^{(k)}      \\ \mbox{{\bf V}}^{(k)} \end{array}
\right) \longrightarrow \left( 
\begin{array}{c}
 \mbox{{\bf V}}^{(k)\star} \\ \mbox{{\bf U}}^{(k)\star} 
\end{array} \right)
\end{equation} 
for the two lowest quasiparticle states $k$=1,2. 
Moreover ({\it ii\/}) we impose a certain, very small, spatial 
cranking frequency
$\hbar\omega_s\sim 0.01$\,MeV to remove the degeneracies in the 
qp spectrum. This does not 
influence the excitation energy,
$\Delta E_{T=1}$, and is further justified  
because the $T$=1 state have, in general, $I\ne 0$.
Finally, ({\it iii\/}) since at the iso-frequency zero the  
alignment $\langle 2qp|\hat t_x | 2qp\rangle_{\omega=0}$=0, we 
determine the cranking frequency $\hbar\omega$ so that our solution
satisfies the condition of $T_x$=$1$.

The results of our calculations are shown in Fig.~\ref{fig2}.
The agreement between the data is more than satisfactory given 
the simplicity of our model. Moreover, because all
parameters follow exactly those
used for the calculations of the $T$=2 states~\cite{[Sat00b]} we have a 
consistent scheme that accounts simultaneously for the
mass excess (the Wigner energy), the $T$=1, and $T$=2 states in
e--e $N$=$Z$ nuclei.

\begin{figure}
\begin{center}
\leavevmode
\epsfysize=8.0cm
\epsfbox{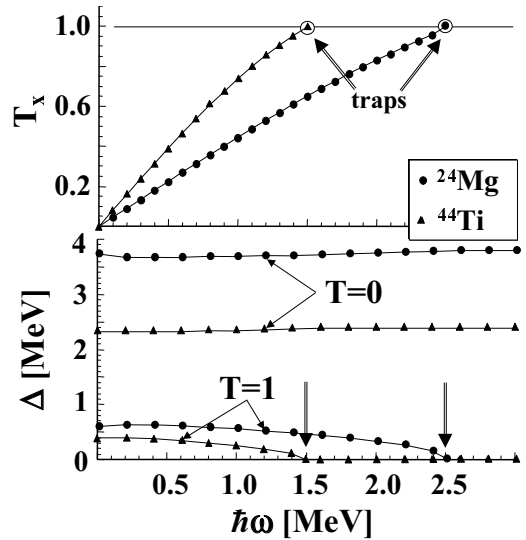}
\end{center}
  \caption{Isospin $T$ ({\bf a}) and isoscalar and isovector pairing gaps
   ({\bf b}) versus rotational frequency. 
   Calculations have been done for $^{24}$Mg and $^{44}$Ti. The figure
   clearly indicate the collapse of isovector pairing correlations giving
   rise to isospin traps.
  }
  \label{fig3}
  \end{figure}

Isoscalar pairing plays a
crucial and, interestingly, different role regarding the nature
of these states. Although the 
$T$=2 states are predicted to be purely isovector paired,
isoscalar correlations are vital 
in restoring the correct inertia parameter~\cite{[Sat00b]} and, hence,
the excitation energy.
In contrast, the $T$=1 states are calculated to have only
{\it isoscalar\/} pairing, see Fig.~\ref{fig3}b. 
Already the 2qp excitation results in strongly reduced
isovector pairing.
With increasing iso-cranking frequency the iso-alignment of the system 
increases smoothly, 
and the weak isovector pairing becomes quenched,
see Fig.~\ref{fig3}. Apparently, by decreasing isovector pairing
the systems gains alignment and, in turn, also  energy.
This process counterbalance the energy loss due to the disappearance
of the isovector pairing.
Once the nucleus reaches the
alignment corresponding to $T_x$=1 it becomes {\it trapped\/}. 
This accounts for the cranking condition of $T_x$=1, since the
2qp configuration decouples from the core and becomes
fully aligned. There is
no collective rotation of the core.
This state does not change until very high iso-frequencies where 
isoscalar pairing is destroyed and either the sp state is reached 
or isovector pairing sets in again. For example, for $^{44}$Ti 
depicted in Fig.~\ref{fig3}, the nucleus is trapped at $\hbar\omega\sim$1.5MeV
and stays there until $\hbar\omega\sim$5.8MeV. 
An interesting consequence emerges from the pairing properties
of this state. Due to isospin symmetry, 
the o--o $T_z$=$\pm$1 states are also expected to have
a predominantly isoscalar pairing field. 
Therefore,  transfer reactions from odd-odd
$T_z$=$\pm$1 to the $T_z$=0,$T$=1 state
and vice versa may be sensitive to
{\sl isoscalar} pairing correlations.

Let us finally turn to the spectrum of o--o $N$=$Z$ nuclei. 
There,
the ground state is determined by the competition between 
the $T$=1 and $T$=0 states, respectively. Though the lowest 
$T$=0 and $T$=1 states are nearly degenerate,  the $T$=0 states
are favored in lighter nuclei (below $f_{7/2}$ sub-shell) whereas the $T$=1 
states become the ground state in heavier nuclei. There are
two exceptions 
from this rule, namely $^{34}$Cl and $^{58}$Cu~\cite{[Zel76]}.
Several authors already pointed out that the structure of
the ground state of o--o $N$=$Z$ nuclei reflects the delicate balance 
between the symmetry energy and 
pairing correlations, and that the energy difference may constitute
a sensitive probe for the role of isovector and isoscalar 
pairing correlations~\cite{[Jan65],[Vog00],[Mac00s]}.
The works of~\cite{[Jan65],[Vog00]} dealt mainly with 
data analysis and did not attempt to provide a microscopic 
explanation. The considerations of Ref.~\cite{[Mac00s]} are based on
the mean-field model with isovector pairing only and  
the  {\it ad hoc\/} assumption that the 
symmetry energy corresponds to $E_{sym}\sim T(T+1)$ 
although it is known that microscopic 
{\it mean-field\/} models yield $E_{sym}\sim (N-Z)^2\sim T^2$.

To better understand the situation in $N$=$Z$ o--o nuclei let us come back 
for a while to the extreme sp picture. In this model two valence 
nucleons can form either an isovector, $T=1$, pair ($|+\rangle|+\rangle$)
giving rise to iso-aligned ground state configuration or 
an isoscalar pair ($|-\rangle|+\rangle$)
forming a  $T$=0 p-h excitation. The energy of both 
states is completely degenerate.
Again, pairing correlations will 
modify this simple picture.

Within the standard mean-field theory for pairing correlations (HFB)
the ground states of o--o $N\ne Z$ nuclei are described as
2qp excitation of the e--e vacuum, 
$\hat \alpha_1^\dagger \hat\alpha_2^\dagger |vac\rangle$.
These ground states are all the 
minimal isospin states $T=|T_z|=|N-Z|/2$. Hence, for reason of
consistency, 
{\sl all} states of minimal isospin in o--o nuclei, including the 
$T$=0 states in
$N$=$Z$ nuclei, have to be treated as 2qp states.  
Therefore, we calculate the $T$=0 ground state of the
o--o $N$=$Z$ nuclei by blocking the 
lowest 2qp states self-consistently, similarly
to the case of e--e $T$=1 states. However,
no iso-cranking is necessary, since $T$=0.

%%%%%%%%%%%%%%%%%%%%%%%%%% figure 3 : TRAPS %%%%%%%%%%%%%%%%%%%%%%%%%%
%%%%%%%%%%%%%%%%%%%%%%%%%%%%%%%%%%%%%%%%%%%%%%%%%%%%%%%%%%%%%%%%%%%%%%%%

\begin{figure}
\begin{center}
\leavevmode
\epsfysize=5.0cm
\epsfbox{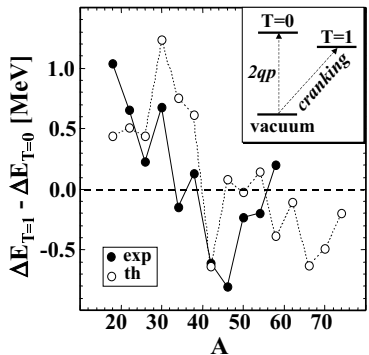}
\end{center}
  \caption{Empirical ($\bullet$) and calculated ($\circ$) excitation 
   energies, $\Delta E= \Delta E_{T=1} - \Delta E_{T=0}$, 
   of the lowest $T$=0 and $T$=1 states in o--o $N$=$Z$ nuclei. 
   The insert indicates schematically the two different excitation
   modes of the $T$=0 and $T$=1 states in our calculations,
   see text for more details.
  }
  \label{fig4}
  \end{figure}

In contrast, due to the isospin symmetry, 
the $T$=1 state of the o--o  $N$=$Z$ nucleus
can be regarded as a linear combination of 
the isobaric analogue states i.e. the ($N$+1,$Z$--1) and ($N$--1,$Z$+1) 
e--e neighbours. 
Hence, it represents the vacuum of an e--e nucleus, however
{\sl excited in isospace}. Since in this case we project on 
good $T_z$=0, the e--e vacuum need to be iso-cranked 
to yield the correct value of $T_x=\sqrt{2}$.  
The difference between the theoretical 
approach to calculate $T$=0 and $T$=1 states
in o--o nuclei is shown schematically in the inset of Fig.~\ref{fig4},
which elucidates the role played by blocking and iso-cranking, 
respectively. 
The difference in structure between these states
is easy to understand qualitatively. Indeed, since no blocking 
but only iso-cranking
is applied for $T$=1 states, isovector pairing is not reduced
at all. However, isoscalar pairing is suppressed
due to the isospin anti-pairing effect~\cite{[Sat00b]}.
In contrast, the $T$=0 state 
experience strongly reduced pairing correlations due 
to the blocking effect. Note, that both isoscalar and isovector 
correlations are reduced as compared to the e--e neighbour.
Hence, the $T$=0 and $T$=1 states in o--o and e--e
$N$=$Z$ nuclei are of different nature,
since they are based on two different fundamental excitations.
It is therefore straightforward to understand the basic differences in
the excitation energy pattern of e--e and o--o nuclei. 
In e--e nuclei, we need to consider both quasi-particle excitations
{\sl and} isospin cranking for the $T$=1 excitation. 
Both are costly in energy and hence the
excitation energy is rather high. 
In o--o nuclei, we simply deal with the competition
between iso-cranking ($T$=1) and 2qp excitation ($T$=0).
Energetically, to first approximation, 
these effects are very similar.

To get a quantitative estimate on the energy difference of the
$T$=0 and $T$=1 states $\Delta E = \Delta E_{T=1} - \Delta E_{T=0}$
in o--o nuclei, we performed
a set of calculations following the rules sketched above. 
The results are presented in Fig.~\ref{fig4}.
As mentioned above, to first order these two basically 
different states are almost degenerate in experiment ($\bullet$). 
Similar result are indeed obtained in our calculations ($\circ$).
This result is particularly interesting because 
it was claimed previously, that this degeneracy is a proof
of lacking $T$=0 pairing correlations~\cite{[Mac00s]}. 
Evidently, these claims were
based upon a poor understanding of the underlying structure of
the elementary excitations allowed in the presence of proton-neutron 
pairing correlations.

%%%%%%%%%%%%%%%%%%%%%%%%%% figure 4 : odd-odd %%%%%%%%%%%%%%%%%%%%%%%%%%
%%%%%%%%%%%%%%%%%%%%%%%%%%%%%%%%%%%%%%%%%%%%%%%%%%%%%%%%%%%%%%%%%%%%%%%%

Note, that in our calculations we obtain not only the near-degeneracy
but also an inversion of the sign of $\Delta E$ which, in agreement with 
experiment, takes place somewhere around the $f_{7/2}$ sub-shell.
The inversion reflects basically the different mass dependence
of the symmetry energy and the pairing correlations.
Since the value of $\Delta E_{T=1}$  is governed
by the symmetry energy, it will decrease with mass as $\sim 1/A$. 
On the other hand, the value of $\Delta E_{T=0}$ is governed by
pairing properties, i.e. depends on the size of
the effective pairing gap including both $T$=0 and $T$=1 pairing
correlations. Apparently, the pairing correlations do not fall off 
with mass as rapidly as $1/A$ giving rise to the inversion.
%-----------------------------  summary  ----------------------------

In summary, we have presented a consistent microscopic explanation
of the pairing phenomena in o--o and e--e $N$=$Z$ nuclei based
on the {\it mean-field\/} approximation.
Our model includes, in a self-consistent manner, both isoscalar and 
isovector pairing correlations,
and takes into account projection onto good particle-number
[within the so called Lipkin-Nogami approximation~\cite{[Sat00a]}],
and isospin [within isospin cranking formalism~\cite{[Sat00b]}].
In e--e $N$=$Z$ nuclei the $T$=1 excitation 
is described as an iso-cranked 2qp configuration. 
According to the model, with increasing iso-frequency,
the valence pair decouples from the fully isoscalar-paired
core and aligns along the x-axis in isospace forming a 
trap at iso-alignment $T_x=1$. 
In o--o $N$=$Z$ nuclei the $T$=1 excitation 
is described by means of the iso-cranked 
o--o 'false vacuum' with $T_x=\sqrt{2}$. 
Hence, this state represents 
a mixture of e--e neighbours ($Z$--1,$N$+1 and 
$Z$+1,$N$--1) in accordance with isospin symmetry. 
The $T$=0 excitations in o--o $N$=$Z$ nuclei, on the other hand,
are treated as 2qp excitations
on top of the o--o 'false vacuum' similar to the standard 
self-consistent BCS treatment of all o--o $N\ne Z$ nuclei.  
The model simultaneously account for  the Wigner energy
the excitation energies of the $T$=1 and $T$=2 states 
in e--e nuclei, the near-degeneracy as well as inversion 
of $T$=0 and $T$=1 states in o--o $N$=$Z$ nuclei.

%\bigskip

This work was supported by the G\"oran Gustafsson Foundation,
the Swedish Natural Science Research council (NFR),
and the Polish Committee for Scientific Research (KBN) under
Contract No. 2~P03B~040~14.

\end{document}